\documentclass{article}
\usepackage{hiph-art}
\usepackage{graphicx}

\newcommand{\fig}[1]{Fig.~#1}
\newcommand{\pc}{$\pi^{\pm}\pi^0$ }
\newcommand{\pn}{$\pi^0\pi^0$ }
\newcommand{\MeV}{\mathrm{MeV} }

\volnumber{19} \issuenumber{1} \edyear{2004}                             
\frompage{000} \topage{000}                                              
\recrevdate{11 February 2005}                                              
\def\rightmark{Low-energy pions in nuclear matter}
\def\leftmark{O. Buss, L. Alvarez-Ruso,  P. Muehlich, U. Mosel, R. Shyam}

\title{Low-energy pions in nuclear matter and $\pi \pi$ photoproduction within a BUU transport model} 
\authors{Oliver Buss \footnote{oliver.buss@theo.physik.uni-giessen.de}
, Luis Alvarez-Ruso, Pascal Muehlich, Ulrich Mosel $^1$ \\Radhey Shyam $^{2}$\\
\index{Buss, 0.} 
\index{Alvarez-Ruso, L.} 
\index{Muehlich, P.} 
\index{Mosel, U.} 
\index{Shyam, R.} 
\\[2.812mm]
{\normalsize
\hspace*{-8pt}$^1$ Institut f\"ur Theoretische Physik, Universit\"at Giessen, Germany\\ 
\hspace*{-8pt}$^2$ Saha Institute of Nuclear Physics, Calcutta, India
}}
 
\abstract{
In the present paper we investigate a method to describe low-energy scattering events of pions and nuclei within a  Boltzmann-Uehling-Uhlenbeck (BUU) transport model. 
Implementing different scenarios of medium modifications, we studied the mean free path of pions in nuclear matter at low momenta and compared pion absorption simulations to data. Pursuing these studies we have shown, that also in a regime of a long pionic wave length the semi-classical BUU model still generates reasonable results. We present results on $\pi$-induced events in the regime of $10\, \MeV\leq T^{\pi}_{kin}\leq 130\, \MeV$ and photo-induced $\pi \pi$ production at incident beam energies of $400-460\, \MeV$. 
}


\makeindex
\begin{document}
\maketitle
\section{Introduction and Motivation}\label{intro}
According to present knowledge, the hadronic world can be described in the terms of Quantum Chromo Dynamics (QCD) which incorporates a spontaneously broken symmetry, the so-called \textit{chiral symmetry}. Since the order parameter $\langle q\bar{q}\rangle$ of this symmetry breaking is expected to decrease by about 30\% already at normal nuclear matter density~\cite{Drukarev:1990kd,Cohen:1992nk,Brockmann:1996iv}, one expects to observe signals for partial chiral symmetry restoration in photonuclear reaction experiments. 

At first glance, the modification of the so-called $\sigma$ or $f_0(600)$ meson inside the nuclear medium promises to be an easily accessible signal for such a partial symmetry restoration. Theoretical predictions expect a shift of its spectral strength to lower masses and a more narrow width due to the onset of the restoration~\cite{Bernard:1987im,Hatsuda:1999kd}. The $\sigma$ meson is very short-lived. Once excited inside a nucleus, it will also very probably decay inside. The excitation of the $\sigma$ meson by a pion or a photon inside the nuclear medium and the later measurement of the invariant mass of its $\pi \pi$ decay products outside the medium should therefore reveal informations about its mass at finite matter density. 

Such experiments have been performed with incident pions by the CHAOS collaboration~\cite{3} and photons in the incoming channel by the TAPS collaboration~\cite{2}. Both experiments have shown an accumulation of strength near the $\pi \pi$ threshold in the decay channel of the $\sigma$. One possible interpretation of this effect is the in-medium modification of this resonance due to partial symmetry restoration. 
In contrast to this we present a possible interpretation based on final-state effects analysing the experiment performed by the TAPS collaboration. 

We need to consider pions with very low energy in nuclear matter to describe the final-state effects of the latter experiments. Since we want to employ a BUU transport model we first need to establish that it makes sense to discuss pions with long wave length in such a semi-classical picture. Actually, this question can only be answered by comparison to quantum calculations and experimental data. Therefore we will validate our model for the treatment of the $\pi \pi$ final state with the description of pion-induced scattering and absorption experiments within the same framework.

Our paper is structured in the following way: In section (\ref{buu}) we briefly introduce our BUU transport model for the treatment of photon- and pion-induced nuclear reactions. In the following section we show results on the mean free path of pions in nuclear matter and discuss consequences of medium modifications of the pion. Hereafter we compare our simulation to experimental results on pion scattering off complex nuclei to validate our model. Finally, in  section (\ref{piPiPhoto}) we conclude with results concerning the $\pi\pi$ photoproduction at incident beam energies of $400-460 \, \MeV$.

\section{The BUU transport model}\label{buu}

\begin{figure}[hb!] 
	\centering
	\includegraphics[height=5cm]{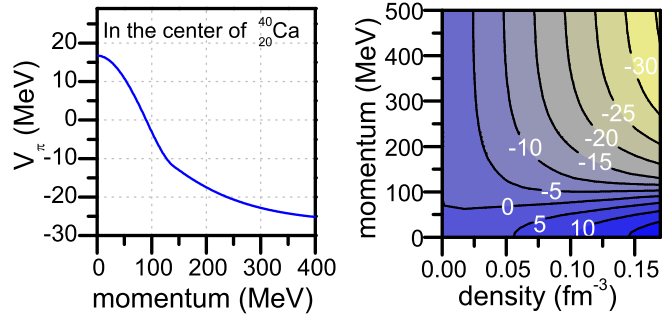}
	\caption{\textbf{Hadronic potential of the pion.} On the left panel we see a plot of the strength of the potential in the center of a $^{40}_{20}Ca$ nucleus as a function of the pion momentum in the nuclear rest frame. The plot to the right shows the overall shape of the potential in units of MeV dependent on momentum and density in symmetric nuclear matter.}
	\label{potPlot}
\end{figure}

Boltzmann-Uehling-Uhlenbeck (BUU) transport models are based on the famous Boltzmann equation, which was modified in the 1930's  by Nordheim, Uehling and Uhlenbeck to incorporate also quantum statistics. It is of semi-classical nature, but can be derived as a gradient-expansion of the fully quantum Kadanoff-Baym-equations. 
In our days it is widely used in heavy-ion collisions, electro- and pion-induced reactions to describe these processes in a coupled- channel approach. Here we want to emphasize on the most important features of the model, for a detailed discussion we refer the reader to~\cite{1,Effenberger:1997rc,Teis:1997kx} and the references therein. 
\subsection{Elementary processes}
The elementary processes of interest are on the one hand side $\pi N \rightarrow \pi N$ and on the other hand side $\pi N N \rightarrow N N$. Both are modeled within our resonance picture, therefore we can express the relevant cross sections as
\begin{eqnarray*}
\sigma_{\pi N \rightarrow \pi N}&=&\sigma_{\pi N \rightarrow R \rightarrow \pi N}+\sigma^{BG}_{\pi N \rightarrow \pi N} \\
\sigma_{\pi N N \rightarrow N N}&=&\sigma_{\pi N N \rightarrow R N \rightarrow N N}+\sigma^{BG}_{\pi N N\rightarrow N N}.
\end{eqnarray*}

The background cross sections denoted by $\sigma^{BG}$ are chosen in such a manner, that the elementary cross section data in the vacuum are reproduced~\cite{1}.

\subsection{Medium modifications}\label{medMod}
As medium modifications we implement besides trivial effects like Pauli-blocking, Fermi-motion of the nucleons and Coulomb forces also different types of hadronic potentials for the particles and density dependend modifications of the decay widths. One needs to emphasize that the imaginary parts of the potentials are not included, since their effect is already accounted for by the collision term.
\paragraph{Potentials}
The potentials are introduced as 0th component of a vector potential in the local rest-frame. For our purposes the most important potentials are those concerning the nucleon, $\Delta$ resonance and the $\pi$ meson. The nucleon potential is described by a momentum-dependend mean-field potential, for explicit details and parameters see~\cite{Teis:1997kx} . Phenomenology tells us that the $\Delta$ potential has to be circa $-30\, \MeV$ deep at $\rho_{0}$. 
Comparing to a momentum independent nucleon potential, which is approximately $50\, \MeV$ strong, the $\Delta$ potential is therefore assumed to be 
\begin{eqnarray*}
	V_{\Delta}=\frac{2}{3}\ V_{nucleon}.
	\label{DeltaPotential}
\end{eqnarray*}
Consequently the $\Delta$ potential has the same shape as the nucleon potential, but is scaled to match the phenomenological value at  $\rho_{0}$. 

For the first time we also implemented a realistic potential for the $\pi$ meson at very low energies. Below $80\, \MeV$ pion momentum a model by Nieves et al.~\cite{osetPot} and above $140\, \MeV$ a first-order $\Delta$-hole result was utilized. Both models were matched in the region of $80-140 \, \MeV$, the result is presented in \fig{\ref{potPlot}}. 

\paragraph{Modifications of the $\Delta$ width}
The $\Delta$ resonance is explicitly propagated in the model. Besides explicit $N\Delta \longrightarrow N \Delta$ collisions the model implements an absorption probability for this resonance according to~\cite{osetWidth} to account for $N\Delta \longrightarrow N N$  and $N N \Delta \longrightarrow N N N$ processes.


\section{Results of the simulations}
\subsection{Pion-induced processes}\label{pionInduced}
Already in earlier works of Salcedo et al.~\cite{salcedoSim}, with only pions as degrees of freedom in the simulation, and of Engel et al.~\cite{engel}, with a precursor version of our present model,  pions with kinetic energies in the regime of $85 - 300 \, \MeV$ have been investigated in transport models. As motivated in the introduction, we now need to investigate even less energetic pions. Therefore we carefully accounted for Coulomb corrections in the initial channel of $\pi$-induced processes and improved the threshold behavior of the cross sections in the model.

\paragraph{The mean free path of pions in nuclear matter}\label{meanFree}
\begin{figure}[ht!]
	\centering
		\includegraphics[width=6cm]{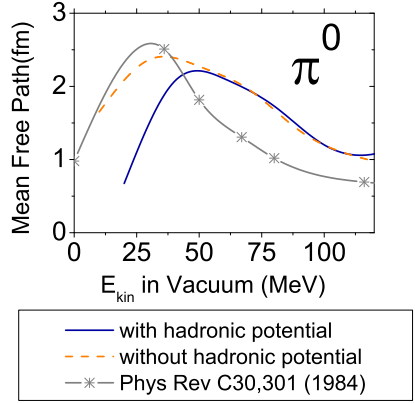}
		\caption{Mean free path of the $\pi^{0}$ meson at normal nuclear matter density. Results of the BUU simulation versus a quantum mechanical analysis carried out by Mehrem, Radi and Rasmussen~\cite{MehremRadi}. The stars denote the quantum mechanical results. The spline connecting the quantum mechanical results is just meant to guide the eye. The two BUU curves differ in the hadronic potential for the pion.}
		\label{OptModelVergleichRami}
\end{figure} 
In an earlier work Mehrem et al.~\cite{MehremRadi} utilized optical model results to estimate the mean free path of the $\pi$ meson in nuclear matter on a fully quantum mechanical basis. 
Their result for the mean free path is shown in \fig{\ref{OptModelVergleichRami}} in comparison to our result obtained with the BUU simulation.

At very low energies the mean free path in the BUU simulation without the hadronic potential for the pion corresponds very well to the result by Mehrem et al.~. 

Including the hadronic potential for the pion we notice that the mean free path goes to zero at low kinetic energies of the pion. There the hadronic potential becomes repulsive. In classical mechanics particles can not propagate in regions where $E_{kin}<V$, this is also true for our BUU simulation which uses classical trajectories for the testparticles. In contrast to classical mechanics, quantum mechanics allows for tunneling into such forbidden regions where $E_{kin}<V$. Therefore the classical interpretation underestimates the mean free path at low energies where the tunneling effect becomes important.

At higher energies Mehrem et al. report a lower mean free path than the results of our simulation. All together the result without the hadronic potential agrees best with the quantum mechanical result. The mean free path is mainly unaffected by the potential above $30  \MeV$ kinetic energy of the pions. Since the mean free path is not directly accessible by experiment, we want to evaluate our model in the context of scattering experiments.

\paragraph{Pion-induced reactions - scattering and absorption}\label{pionAbs}
\begin{figure}[ht!]
\label{absPlot1}
\begin{center}
\includegraphics[height=135mm]{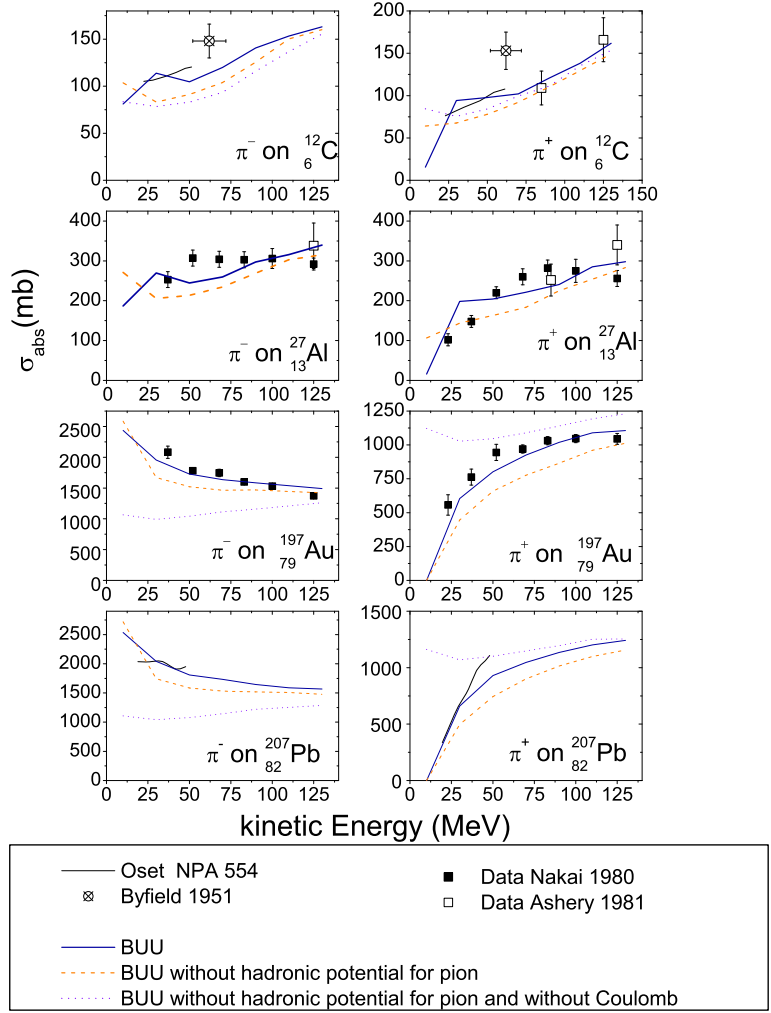}
\caption{Pion absorption on nuclei depending on the choice of potentials for the pion. Data points are taken out of~\cite{byfield},\cite{ashery} and~\cite{nakai}. The optical model results of \cite{osetPot}, denoted 'Oset NPA 554', are shown for comparison.}
\end{center}
\end{figure}
Results on pion absorption for various nuclei are presented in \fig{3}. Actually, the overall agreement to the data is very satisfying. Considering the fact that the pions have very long wave lengths at such low energies, the success of our semi-classical BUU model is quite astonishing. Due to the long wave length one expects also multi-particle collisions to be important, but actually we have only $2\rightarrow 2$ and $2\leftrightarrow 3$ processes implemented explicitly in the simulation.

The overall agreement to the data is much better with the hadronic potential included. Introducing this potential, one accounts also for multi-particle interactions for the pion which are effectively incorporated in this mean field potential. The effect of the potential on the trajectories of the pions is more important than its effect on the mean free path. We see that also above $T_{kin}=30\, \MeV$, where the mean free path is unaffected, the bending of the trajectories due the attractive potential causes higher cross sections.
One has to conclude that the mean field of the pion is an important ingredient for transport simulations at such low energies and that our transport model is well suited to describe low energetic pions down to $T_{kin}\simeq 20\, \MeV$. Nevertheless one must be cautious when the kinetic energy of the pion drops below the value of its potential. 
\subsection{Photon-induced $\pi \pi$ production in nuclei}\label{piPiPhoto}
\begin{figure}[ht!]
\begin{center}

\includegraphics[height=75mm]{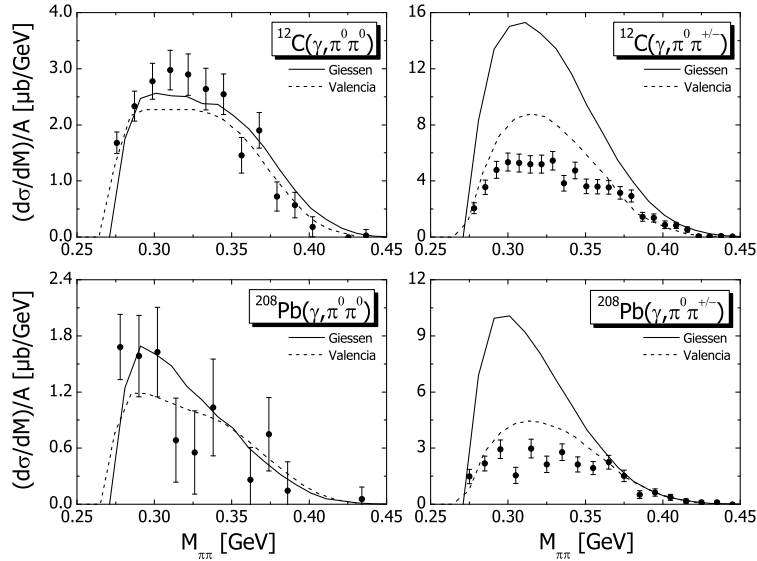}

\caption{Two pion invariant mass distributions for \pn and \pc photoproduction of $^{12}$C and $^{208}$Pb. The continuous lines labeled "Giessen" represent the results of~\cite{TwoPiPaper} and the dashed lines labeled "Valencia" depict the results of Ref.~\cite{Roca}. The experimental data stem from Ref.~\cite{2}.}
\label{twoPi}
\end{center}
\end{figure}
In the previous section it was important to see that a transport approach is still meaningful in the regime of long pionic wave length. In~\cite{TwoPiPaper} we presented in detail results on the double-pion photoproduction off nuclei utilizing this transport approach. There, the treatment of pion absorption differs from the treatment in the last sections. In oder to compare to a work by Roca et al.~\cite{Roca} we there implemented the same absorption probability for the pion as in their work. Therefore we did not propagate the resonances in our model, but rather used experimental data to describe the quasi-elastic $N\pi \rightarrow N \pi$ scattering and used the method of~\cite{Roca} to model pion absorption. A publication with the full BUU final state is in preparation.

Already in~\cite{TwoPiPaper} we indicated the relevant role of some conventional final-state interaction (FSI) effects in two-pion photoproduction in nuclei. In \fig{\ref{twoPi}} we compare to the model by Roca et al.~\cite{Roca}, which uses a purely absorptive final-state prescription and does not include elastic scattering and charge exchange processes for the pions. 

Elastic scattering of the pions with nucleons of the target nucleus yields a loss of kinetic energy of the scattered pions. Therefore, in average, it leads to a reduced invariant mass of the pion pair. Charge exchange reactions produce a non-negligible contribution to the $\pi\pi$ mass spectra, giving rise to an even more pronounced accumulation of spectral strength in the low invariant mass region. On the other hand, side-feeding especially from the $\gamma N\to\pi^+\pi^-N$ reaction followed by a $\pi N$ collision with charge exchange increases the differential cross sections in the \pc channel to values significantly above the data. This effect cannot be accounted for in models involving purely absorptive FSI. The apparently stronger absorption of pions in the \pc channel as compared to the \pn channel is still an unsolved problem both from the experimental and theoretical point of view. 

However, our explanation of the observed downward shift of the invariant mass spectrum follows completely well established nuclear physics effects without modifications of the initial $2\pi$ production process, as one would need to do in the scenario of a in-medium modification of the $\sigma$ meson. Even if a complete understanding of $\pi\pi$ photoproduction off complex nuclei is not yet possible, these effects have to be accounted for in any serious calculation. We therefore conclude that, given our present understanding, the observed target mass dependence of the \pn invariant mass spectrum does not provide an unmistakable evidence for the modification of the $\pi\pi$ interaction or, respectively, for the properties of the $\sigma$ meson at finite nuclear densities. \section*{Acknowledgments}
The authors acknowledge valuable discussions with A. Larionov, J. Lehr and M. Post. This work was supported by DFG. One of us, LAR, has been supported by the Alexander von Humboldt Foundation. 

\vfill\eject
\end{document}